\documentclass[fleqn,10pt]{article}
\usepackage[utf8]{inputenc}
\usepackage[T1]{fontenc}
\usepackage{lineno}
\usepackage{authblk}
\usepackage{hyperref}
\usepackage{graphicx}
\usepackage{bm}% bold math
\usepackage{amsmath,amssymb}
\usepackage[left=2cm,right=2cm,top=2cm,bottom=2cm]{geometry}

\newcommand{\eg}{{\it e.g.}, }
\newcommand{\ie}{{\it i.e.}, }

\newcommand{\fsub}[1]{$_\mathrm{#1}$}

\title{Density-of-states similarity descriptor for unsupervised learning from materials data}

\author[1, *]{Martin Kuban}
\author[1]{Santiago Rigamonti}
\author[1]{Markus Scheidgen}
\author[1]{Claudia Draxl}
\affil[1]{ Humboldt-Universit{\"a}t zu Berlin,Institut f\"ur Physik und IRIS Adlershof, Berlin, 12489, Germany}

\affil[*]{corresponding author: Martin Kuban (kuban@physik.hu-berlin.de)}

\begin{document}
\maketitle

\begin{abstract}
We develop a materials descriptor based on the electronic density of states and investigate the similarity of materials based on it. As an application example, we study the  Computational 2D Materials Database that hosts thousands of two-dimensional materials with their properties calculated by density-functional theory. Combining our descriptor with a clustering algorithm, we identify groups of materials with similar electronic structure. We characterize these clusters in terms of their crystal structure, their atomic composition, and the respective electronic configurations to rationalize the found (dis)similarities. 
\end{abstract}

\flushbottom

\thispagestyle{empty}

\section*{Introduction}
The creation of databases for computational materials science has led to a huge amount of stored calculations, exceeding by far any human's ability to comprehend the information in it. Thus, algorithmic data-analysis methods need to be leveraged to allow knowledge extraction from this large pool of data. 
Domain-specific search interfaces, provided by public databases \cite{NOMAD, Haastrup_2018, aflowlib, materials_project, OQMD}, are one way to make information findable. These interfaces allow researchers to identify materials of their interest, \eg in terms of structural features like space group or atom types, or in terms of properties like the electronic band gap. However, such features provide little insight only.
Furthermore, the use of search interfaces is limited to mostly confirmatory analysis: Having a concrete physical mechanism in mind, \eg the change of properties of alloys with stoichiometry, researchers can manually search materials that allow to confirm, or deny, a hypothesis.

Learning from data, however, is not limited to this kind of analysis. 
For instance, relations between materials in terms of certain properties, can become (only) apparent in large quantities of data. 
To reveal such relations and make use of them, both in-depth understanding of when we consider materials to be similar as well as powerful data-analysis methods are required.
A prerequisite for understanding how different materials relate to one another is the availability of descriptive, numerical representations (descriptors), that accurately capture (dis)similarities, \eg stemming from the atomic and/or electronic structure.

In the past years, several descriptors of the atomic structure have been published \cite{SOAP, symmetry_functions, MBTR} and successfully applied for the prediction of material properties using machine learning (ML) techniques. However, descriptors based on the electronic structure are not well established in the ML community. In early work of Isayev and coworkers \cite{Curtarolo1}, descriptors of both the electronic density-of-states (DOS) and the band structure are used to create a graphical representation of more than 20000 materials from the AFLOWlib database. More recently, supervised ML was proposed \cite{learningDOS} to predict electronic densities-of-states by their decomposition in local atomic contributions. Furthermore, a descriptor based on atomic distances, the projected densities of states (PDOS), and the Kohn-Sham band-gap was shown \cite{C2DBprogress} to improve the prediction of computationally expensive material properties.

The majority of ML approaches in materials science focus on speeding up research. This concerns, for instance, the prediction of materials properties that are time-consuming to compute, like the electronic band gap, or the optimization of established methods, \eg speeding up molecular-dynamics simulations through ML-based force fields. Thereby, highly non-linear ML models and/or complex material descriptors are necessary to achieve decent accuracy of predictions. Moreover, the underlying data are typically considered only as input for the ML models, and are not further analyzed.

In this work, we aim at obtaining deeper understanding of large materials data spaces by rationalizing the reasons behind features that materials may share. We demonstrate our approach by the similarity of materials in terms of their electronic properties. To this extent we develop a tunable DOS fingerprint that encodes the DOS of a material into a binary-valued two dimensional (2D) map, stimulated by the work of Ref. [\cite{Curtarolo1}]. Combining it with unsupervised ML methods, we showcase its use by revealing similarities in the electronic structure of materials from the Computational 2D Materials Database (C2DB)\cite{Haastrup_2018}. We are able to uncover not only expected trends, \eg clusters consisting of materials containing isoelectronic substitutions of atomic species, but also unexpected correlations, \eg clusters of structurally very different materials. Our results show that \textit{explorative} analysis of a database allows for finding relations between materials which could not be foreseen without comprehensive, data-driven analysis.

\section*{Results \label{sec:results}}

\subsection*{Clustering}

\begin{figure}[htb]
    \centering
    \includegraphics[width = 0.45\textwidth]{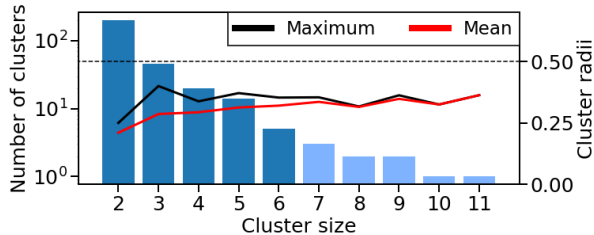}
    \caption{Distribution of cluster sizes (blue bars) and maximal (black line) and mean (red line) cluster radii of the DOS clusters in the data set using a similarity threshold of $S_\mathrm{min} = 0.75$. The dashed line indicates the maximal possible cluster radius for this threshold. The bars in light blue indicate the clusters that are used to generate the similarity matrix in Fig. \ref{fig:DOS_simat}.}
    \label{fig:cluster_dist}
\end{figure}

To identify sets of similar materials, we use the clustering algorithm defined in the methods section. Employing a similarity threshold of $S_\mathrm{thres} = 0.75$, we find $294$ distinct clusters that contain in total $\sim$ $23\%$ of the materials in the entire data set. The remaining $2697$ \textit{orphans} are less similar than $S_\mathrm{thres}$ to any other material in the data set and are not further considered in this specific analysis presented here. We call the materials in a cluster its \textit{members} and identify the \textit{size} of the clusters as the number of its members. The compactness of a cluster is determined by its \textit{radius} $r_c = 1 - S_\mathrm{min}$, with $S_\mathrm{min}$ the minimum similarity between any two members of the cluster. Figure \ref{fig:cluster_dist} presents the distribution of clusters sizes on a logarithmic scale  together with the maximal and mean cluster radii for clusters of a given size. About two third ($200$) of the clusters contain only two materials. Since the clustering algorithm requires that any member of the cluster has a similarity of $S_\mathrm{thres}$ to the reference material, the cluster radii for two-point clusters are as low as $r_c \leq 0.25$. The mean cluster radii for the clusters with more than two members increase to $r_c \sim 0.4$ with increasing cluster size. Interestingly, even though the clustering algorithm allows for the maximal cluster radius to be as large as $r_c = 0.5$, the maximal cluster radii of the discovered clusters are all smaller than 0.4.
We note that the here chosen parameters serve the purpose of showcasing our approach. Both, the similarity threshold as well as the energy range can be varied to focus such analysis on certain aspects of the data. For instance, to search only for compact clusters, the minimal similarity threshold can be increased. This, however, reduces the number of discovered clusters and their size, which ultimately prevents the discovery of meaningful clusters. Conversely, the reduction of the similarity threshold increases both cluster size and number of clusters, to the expense of larger cluster radii. Too large cluster radii bare the risk of masking meaningful relations between data points in large clusters that hinder the automatic analysis of clusters.
\begin{figure}[tb]
    \centering
    \includegraphics[width = 0.45\textwidth]{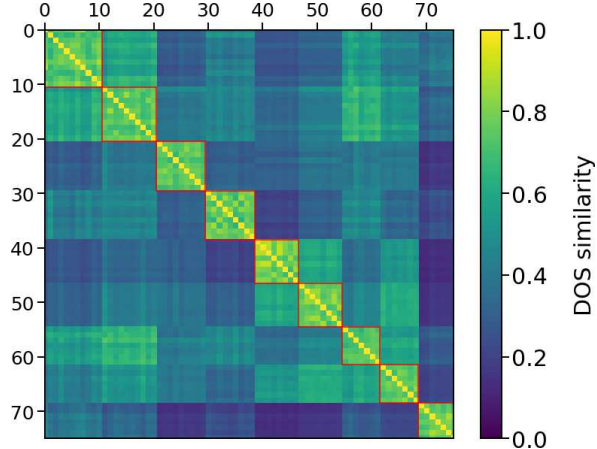}
    \caption{Similarity matrix for materials in clusters with more than six members. The red boxes indicate the clusters detected by our algorithm.}
    \label{fig:DOS_simat}
\end{figure}
To illustrate the similarity relations between materials, we calculate pairwise similarities between all materials in the data set, \ie a symmetric matrix with elements $S_{ij}=S(\bm{f}_i, \bm{f}_j)$. In other words, each column and row of this matrix corresponds to the similarities of a single material to the rest of the data set. The diagonal elements of the matrix are identical, $S_{ii} = 1$, as they describe the similarity of each material with itself. An excerpt of the full matrix can be seen in Fig. \ref{fig:DOS_simat}. The order of columns and rows has been chosen according to the cluster sizes, \ie such that the largest cluster appears in the top left corner of the matrix, and the cluster radius decreases with increasing index. The color code makes apparent that many of the clusters are very dissimilar to each other, \ie the average similarity of the cluster members to the rest of the data set is low. For some others, however, the opposite is the case, and one could expect that, choosing a smaller threshold would merge them, as pointed out above. An example for this, will be given in the following section.
\subsection*{Analysis of selected clusters}
In the following we analyze individual clusters and reason why the materials in these clusters are similar.
\begin{figure}[tb]
    \centering
     \includegraphics[width = 0.45\textwidth]{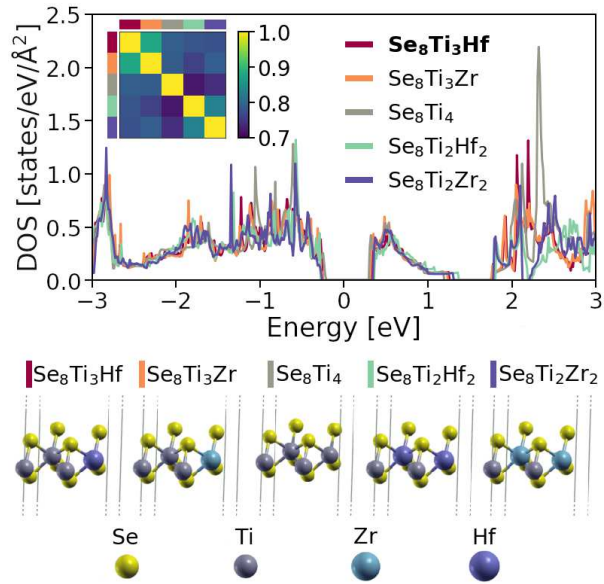}
    \caption{Densities of states (top) and unit cells (bottom) of the materials of a selected DOS cluster. The Fermi level is located at $E = 0$ eV. The cluster center, Se\fsub{8}Ti\fsub{3}Hf, is indicated in bold font in the legend. The inset shows their similarity matrix, where the color code is adapted to reflect the high similarities between the cluster members. Here, sub-clusters become visible for materials with the same number of substituents, \ie the materials containing one or two substituents are more similar to one another than to the other cluster members.}
    \label{fig:HfZrTiSe_cluster}
\end{figure}

\subsubsection*{Isoelectronic substitutions}
\label{sec:iso}
Figure \ref{fig:HfZrTiSe_cluster} presents the DOS and crystal structures of five transition-metal dichalcogenides (TMDC) forming a cluster. Its cluster radius $r_{c}=0.28$ is close to the mean for this cluster size (see Fig. \ref{fig:cluster_dist}). Visual inspection of the corresponding DOS reveals a pronounced overall similarity in terms of \textit{i)} the shape of spectra inside the feature region $|E|\lesssim 2$eV, and  \textit{ii)} the size of the PBE band-gap that varies from 0.52\,eV (TiSe\fsub{2}) to 0.65\,eV (Hf\fsub{2}Ti\fsub{2}Se\fsub{8}). Above 2\,eV, the DOS become more dissimilar, as expected from the coarser representation of the DOS outside the feature region.

Considering the crystal lattice, the cluster members are very similar. All materials consist of a layer of TMs between two layers of Se. The cluster contains the binary phase (TiSe\fsub{2}), as well as ternary phases, where either one or two Ti atoms are substituted with either Hf or Zr or both. The latter type has only minor influence on the DOS. This does not come to a surprise as all substitutions within this cluster are isoelectronic, \ie with atomic species from group 4 of the periodic table of elements (PTE). We note here that there exists another cluster of materials with the same structural prototype, containing, among other materials, 
the binary compounds Se\fsub{8}Hf\fsub{4} and Se\fsub{8}Zr\fsub{4}. These materials form a separate cluster because they have higher PBE band gaps, ranging between 0.72\,eV (Se\fsub{8}TiHf\fsub{3}) and 0.82\,eV (Se\fsub{8}HfZr\fsub{3}). Choosing a lower similarity threshold, these clusters merge.

\begin{figure}[htb]
    \centering
    \includegraphics[width = 0.35\textwidth]{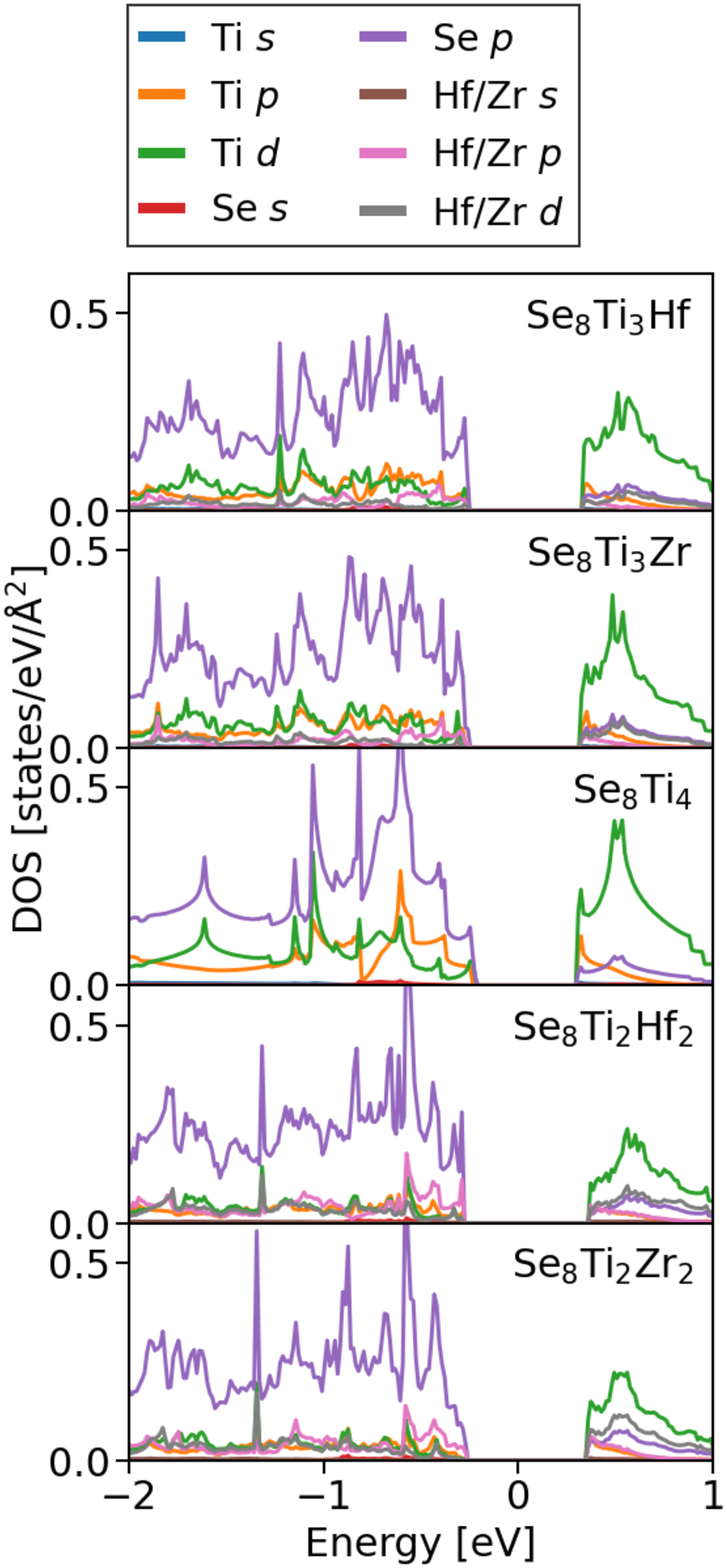}
    \caption{PDOS of the materials in the cluster shown in Fig. \ref{fig:HfZrTiSe_cluster}. The Fermi level is located at $E = 0$ eV. Although the contributions of individual orbitals vary between the different materials, due to their similar shape, their sum, \ie the total DOS, is similar.} 
    \label{fig:HfZrTiSe_PDOS}
\end{figure}

To further demonstrate the isoelectonic behavior of the materials of the here considered cluster, we compare their PDOS in Fig. \ref{fig:HfZrTiSe_PDOS}. Their valence bands are mainly composed of fully occupied Se $4p$ states. The conduction bands have predominant Ti $3d$ character, with additional contributions from $4d$ or $5d$ states of Zr and Hf (when present)\cite{Pal2019}. The latter lie all in the same energy range and sum up to the same number of $d$ states of the four group-IV TMs. The hybridization of TM-$3d$ with Se-$p$ orbitals is evident from small contributions of $d$ states in the valence region and Se-$p$ states in the conduction region. In sum, the similarity of the electronic spectrum of these materials becomes clear: The replacement of Ti by either Hf or Zr does not alter the valence band, while the conduction states are composed of empty $d$ shells of the transition metals, which amount to the same number of empty states.

Several other clusters exist in the data set which consist of isoelectronic materials, \ie they may contain different elements but have the same number of valence electrons. To discover them, we make use of the PTE descriptor introduced in the methods section. Overall, our descriptor identifies $230$ clusters each of them having the same $\overline{c}_{m}$ for all its members (compare Sec. Methods, Eq. \ref{eq:average_column_descriptor} for details).
This number of clusters corresponds to 78\% of all clusters, and 16.5\% of materials in the full data set. Therefore we conclude that isoelectronicity is a main reason for the similarity of the DOS of the materials in the C2DB. 88.8\% of all clusters contain at least two materials that have the same $\overline{c}_{m}$.

\subsubsection*{Materials with isoelectronic surface groups}
\label{sec:iso_surface}

\begin{figure}[htb]
    \centering
    \includegraphics[width = 0.45\textwidth]{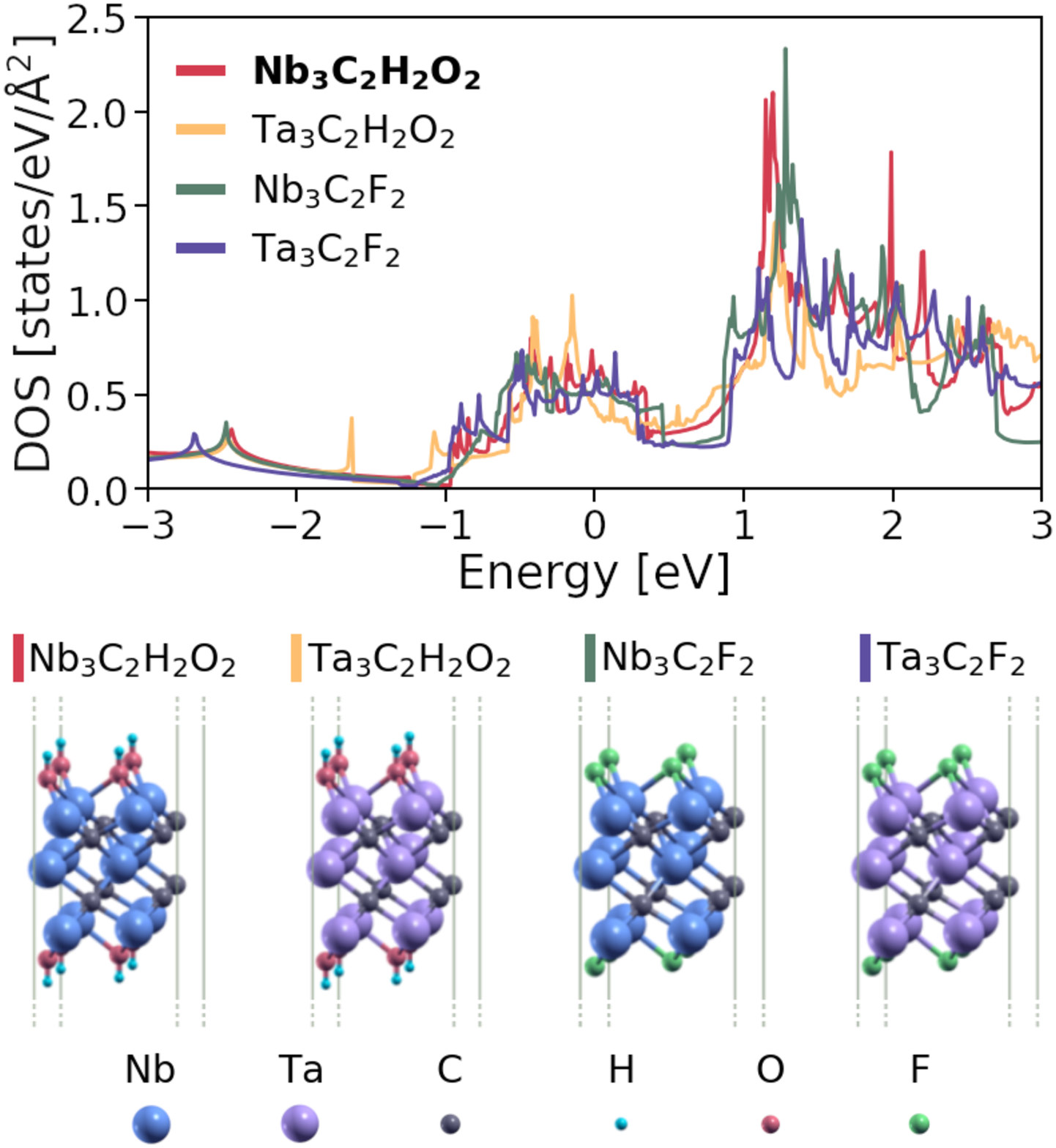}
    \caption{DOS (top) and atomic structures (bottom) of materials with isoelectronic surface groups. The Fermi level is located at $E = 0$ eV. The cluster center is indicated in bold face. For increased visibility the unit cell is repeated in both in-plane directions.}
    \label{fig:F-OH_cluster}
\end{figure}

The second most common origin of similarity concerns the substitution of flourine atoms at the materials' surfaces with OH groups. Figure \ref{fig:F-OH_cluster} presents an example of such clusters. These metallic materials consist of five alternating layers of carbon and either Ta or Nb. Again, Ta, and Nb are isoelectronic, \ie from group 5 of the PTE. At the two surfaces of the materials, either F atoms or OH groups form bonds with the underlying TM atoms.

The cluster radius is $r_{c} \sim {0.28}$, which is close to the mean value for four-point clusters. The general shapes of the curves are similar. Inspection of the PDOS (not shown), reveals that the whole spectrum is dominated by $d$ states of the TM. They hybridize with C and TM $p$ states. The $p$ bands have the largest contribution around $E\simeq0$ and $E\simeq1.5$, where also significant contributions from F and O $p$ states are present. Thus, the saturated O in the hydroxyl group acquires an electronic configuration analogous to F and binds similarly to the TM atoms. In other words, the OH group can be regarded as isoelectronic to F. Minor differences between the spectra, as for instance displaced peaks below -1 eV, mainly originate from multiple van Hove singularities which are very sensitive to the precise location of band extrema and flat bands.

In total, we find $33$ clusters in the data set where F at the surface is interchanged with an OH group. In most of these cases, these clusters contain also sets of materials with other isoelectronic substitutions. Let us note in passing that, despite the fact that the similar behavior of flourine atoms and  hydroxyl groups is well-established expert knowledge in chemistry and electronic-structure theory, it is not trivial to access such knowledge in an automatized manner. So far, the search interfaces of most databases, as well as descriptors for machine learning, rely on structural features, \eg the chemical formula or the number of atoms in the unit cell. Thus, similar materials with, \eg different numbers of atoms in the unit cell are unlikely to be found by these methods.

\subsubsection*{Role of crystal lattice}
\begin{figure}[htb]
    \centering
    \includegraphics[width = 0.45\textwidth]{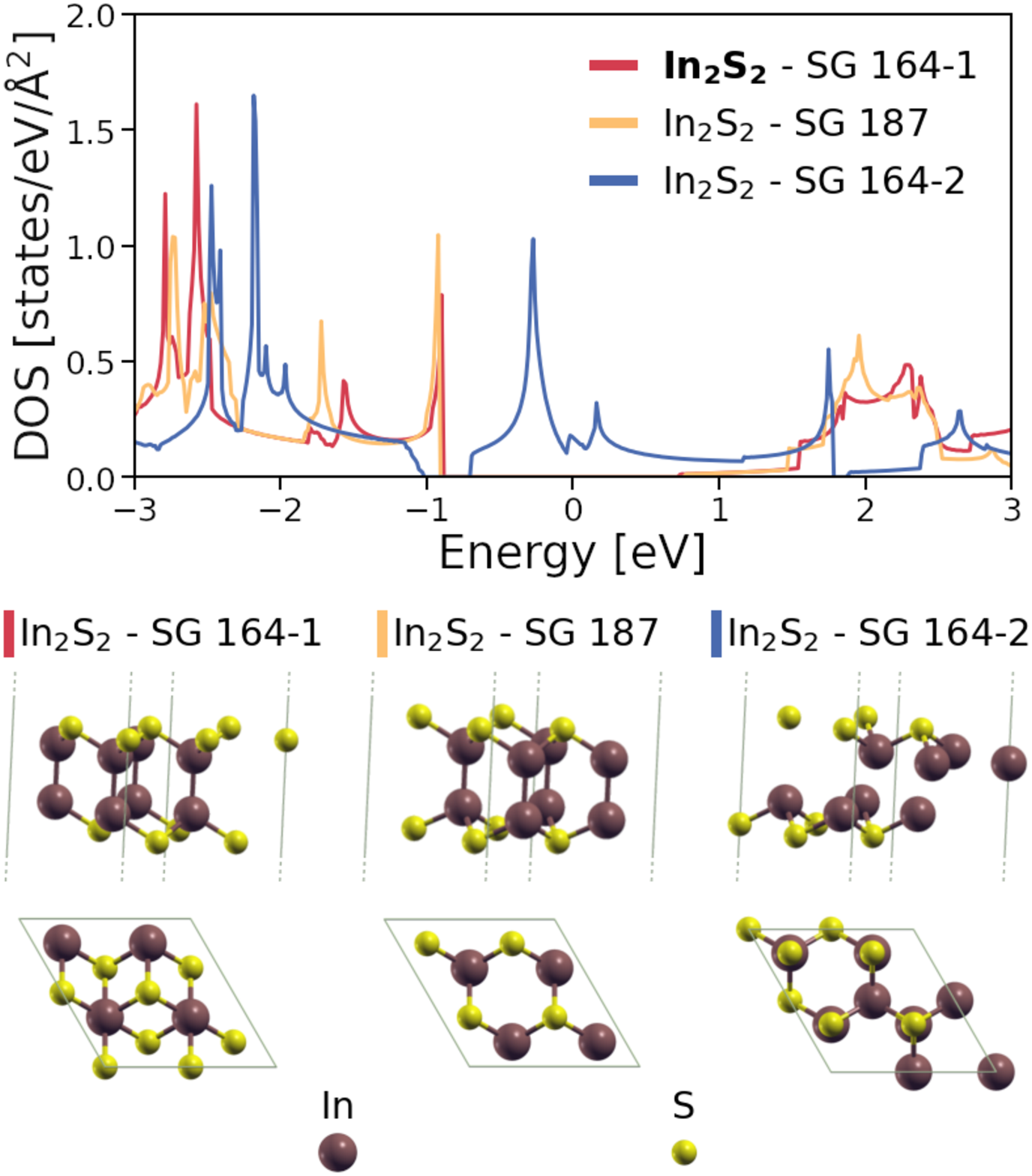}
     \caption{DOS (top) and atomic structures (bottom) of In\fsub{2}S\fsub{2}. The Fermi level is located at $E = 0$ eV. The structures with SG 164-1 and SG 187 form a DOS cluster. The structure with SG 164-2 (right) has a dissimilar DOS and is not part of the cluster. The unit cells are repeated in both in-plane directions to increase visibility.}
    \label{fig:InS_cluster}
\end{figure}

Now, we focus our search on clusters of materials composed of identical atoms but different host lattices. Figure \ref{fig:InS_cluster} presents the data from three different phases of In\fsub{2}S\fsub{2}, which belongs to the class of post-transition metal chalcogenides. We designate them by their SG (value of their SG descriptor, \textit{cf.} Methods), where we distinguish the two materials with SG 164  as SG 164-1 and SG 164-2. The semiconducting structures resulting from the phases SG 164-1 and SG 187 form a cluster due to the similarity of their DOS throughout the whole observed energy range. For comparison, we show the DOS of a third phase, SG 164-2, that shares the symmetry with the first material, however, shows markedly different behavior and, thus, is not part of the same cluster. 
While the similarity coefficient between the clustered materials is 0.76, the corresponding values with the third phase are 0.32 and 0.34, respectively. This finding goes hand in hand with the fact that the clustered materials have medium-sized band gaps of $1.60$ (SG 187) and $1.68$ eV (SG 164-1), respectively \cite{c2dburl}, the third one is metallic. Despite sharing the space group, the two phases with SG 164 show significant structural differences as evident from the top views of the unit cells depicted in Fig. \ref{fig:InS_cluster}. This can be further illustrated considering the stacking of In and S layers: For SG 164-1, the layer sequence corresponds to ABBC stacking; for SG 187, it is ABBA; for SG 164-2, it is ABDC. The (dis-)similarity of the different phases lies in the particular electronic configuration acquired by the atomic species: In the semiconducting phases, In adopts covalent bonding, manifested by a valence band that is dominated by hybridized S and In $p$ states\cite{Zhuang2013}. In this electronic configuration, the In atoms are tetrahedrally coordinated by three S atoms and one In atom. Here the In-In bond length is $d_\mathrm{In-In}$=2.82\;\AA. In the metalic phase, In atoms form metallic bonds with a significant contribution from In $s$ states. In this case, every In atom is coordinated with three In atoms, and the bond length is $d_\mathrm{In-In}$=3.62\;\AA, which is close to that of bulk metallic In (3.38\;\AA). The metallic phase is metastable as compared to the semiconducting ones\cite{c2dburl}.

We note that in this case the dissimilarity between the semiconducting and metallic phases can neither be explained by either the SG nor the PTA descriptors. Nonetheless, our DOS similarity search is able to capture the underlying electronic configuration and put together structures with identical atomic coordination, albeit different crystal structure.

\subsubsection*{Outliers}

\begin{figure}[hbt]
    \centering
    \includegraphics[width = 0.45\textwidth]{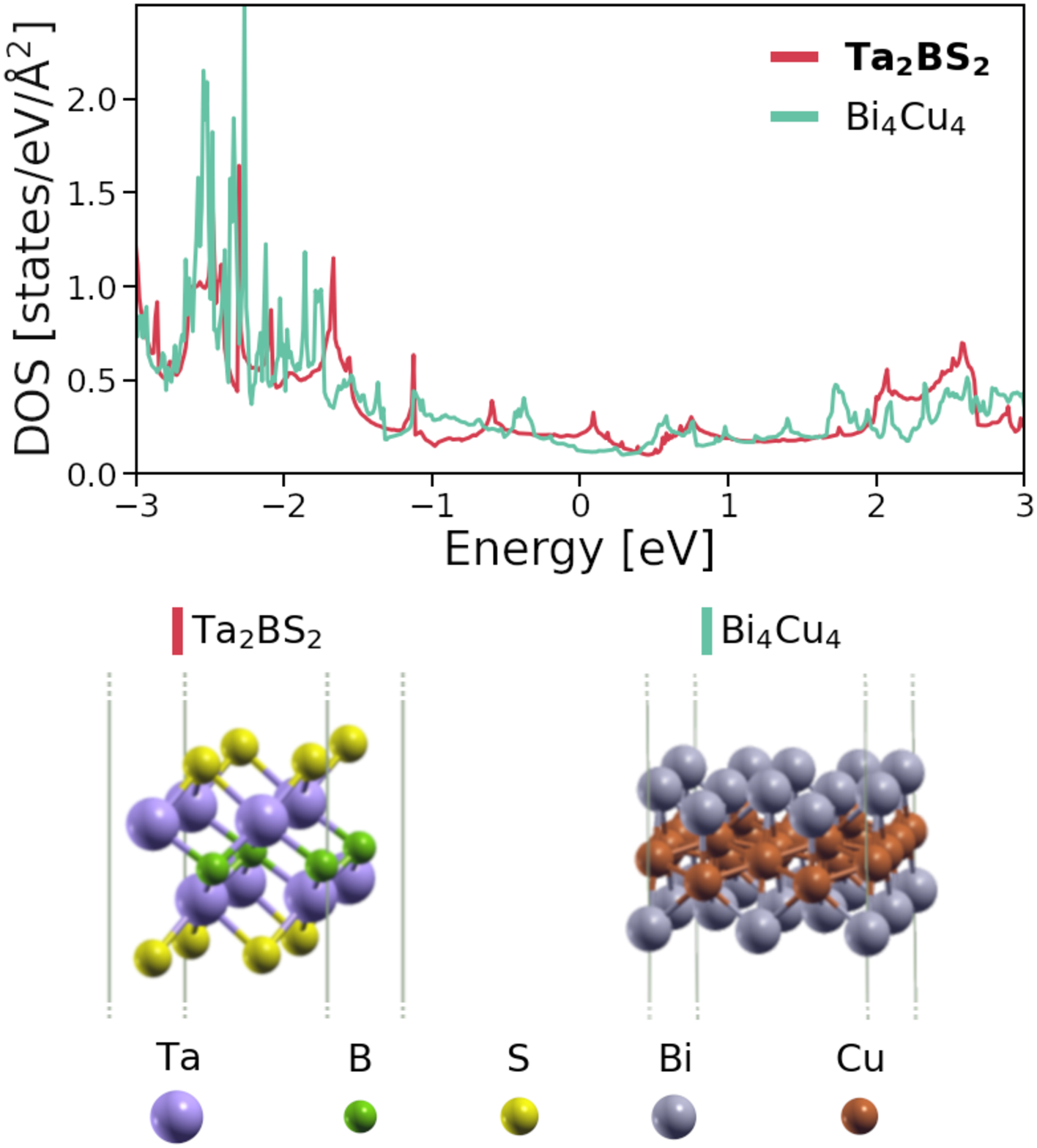}
    \caption{DOS (top) and atomic structures (bottom) of a cluster of materials that neither share atomic species not crystal structure. To increase visibility, the unit cell is repeated in both in-plane directions. The Fermi level is located at $E = 0$ eV.}
    \label{fig:unexp_cluster}
\end{figure}
\begin{figure}[hbt]
    \centering
    \includegraphics[width = \linewidth]{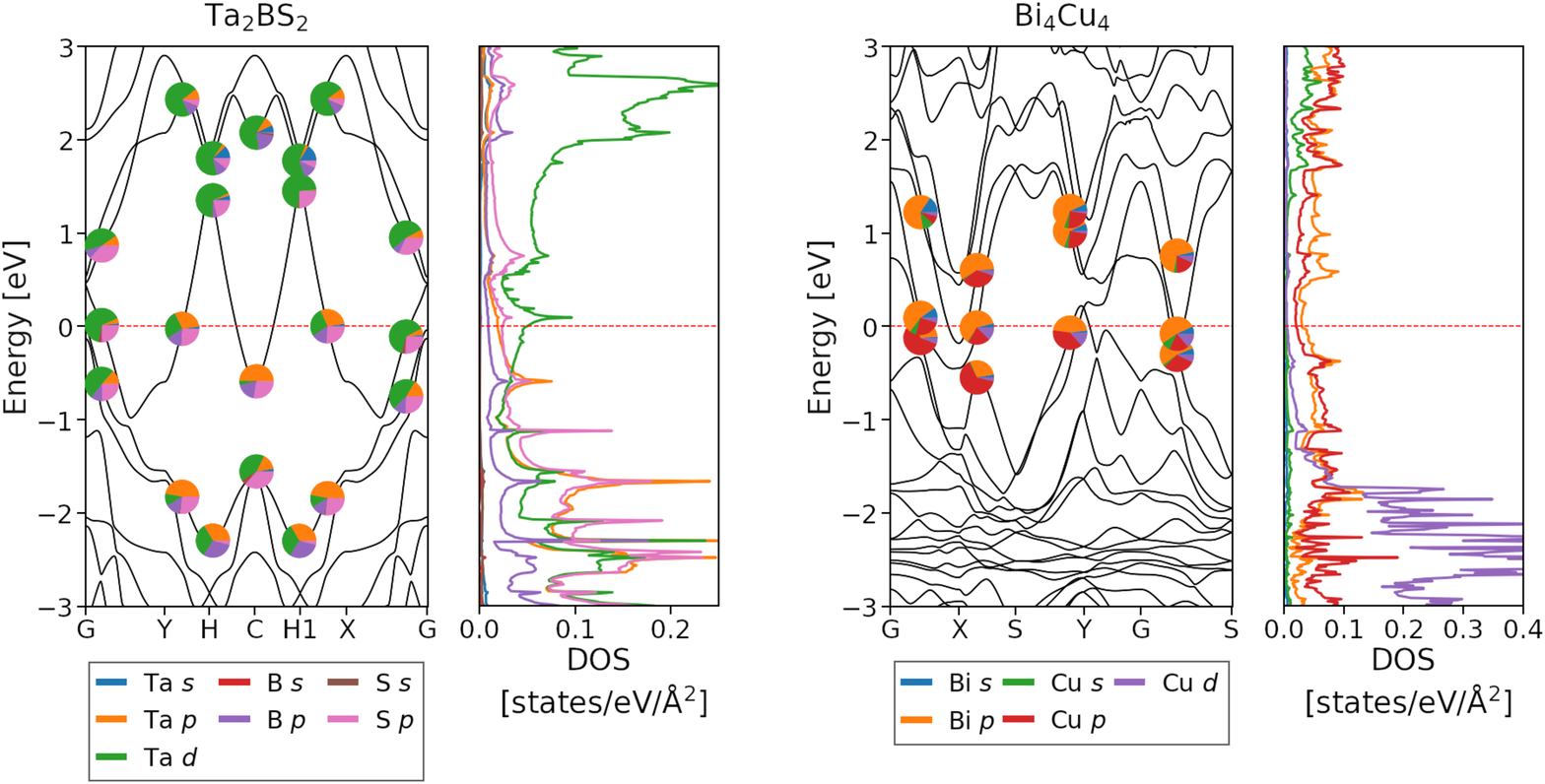}
    \caption{Band structures and projected DOS of Ta\fsub{2}BS\fsub{2} (left) and Bi\fsub{4}Cu\fsub{4} (right), indicating the atomic characters. The Fermi energy is located at E = 0.}
    \label{fig:outlier_bs}
\end{figure}

Overall, there are 25 clusters in the data set with materials that are neither isoelectronic nor share the crystal lattice, \textit{i.e.} they cannot be explained by our SG and PTE descriptors. Therefore, as a final example, we focus on a cluster that consists of two materials that have no apparent similarities in their atomic structures, neither in symmetry nor in composition. They are presented in Fig. \ref{fig:unexp_cluster}. While Ta\fsub{2}BS\fsub{2} has the trigonal space group 164, Bi\fsub{4}Cu\fsub{4} is characterized by an orthorhombic lattice with SG 51. Unaffected by their structural dissimilarities, the DOS of both materials resemble each other with a similarity coefficient of 0.76, \ie slightly above the threshold of $S_\mathrm{min} = 0.75$. Both materials exhibit a nearly constant DOS between $-1.5$  and 2\,eV, while it increases below. To get a deeper insight, we show in Fig.~\ref{fig:outlier_bs} the band structures of Ta\fsub{2}BS\fsub{2} and Bi\fsub{4}Cu\fsub{4}, indicating the atomic character of the bands, together with the corresponding DOS projected on the different atomic species. While in the case of Ta\fsub{2}BS\fsub{2}, only one band with mixed $p$-$d$ character crosses the Fermi level, the energy spectrum of Bi\fsub{4}Cu\fsub{4} exhibits several bands composed almost exclusively of Bi and Cu $p$ states, giving rise to a more complicated topology of the Fermi surface. We conclude that the similarity of these materials' DOS is {\it accidental} and this cluster can be indeed be considered as an outlier.

\section*{Discussion}

In this work, we have presented a fingerprint of the electronic DOS that allows one to quantitatively evaluate the similarity of materials in terms of their electronic structure. We have applied this fingerprint to the C2DB database, a large, heterogeneous data-set of two dimensional materials. Based on our similarity measure, we have devised a clustering algorithm to filter the data for sets of materials that exhibit pronounced similarities to one another. A significant number of (small) clusters have been identified and further analyzed. More specifically, 23\% of the materials can be associated with at least one other material in the data set. The majority of similarities in these particular materials can be explained by the similarity of the valence configuration of the involved atomic species, thus confirming physical expectations. This \textit{confirmatory} analysis has been performed in an automatic fashion based on descriptors that are able to identify those physical reasons. In this way, we could identify, for instance, 16.5\% of materials being isoelectronic to at least another material of the database and exhibiting a similar DOS. Our approach could be easily extended by introducing new descriptors with the potential of explaining other reasons for materials to be similar.

Summarizing, our method provides a means of analyzing large data sets from electronic-structure theory and contributes to understanding, controlling, and selecting such data in view of their re-use in other contexts. Last but not least, the finding of accidental -- unexpected -- similarities may be of relevance in technological applications, where considering materials with different composition but similar properties could lead to \eg structures that are easier to synthesize or reveal other properties that are superior to those of the known materials.

\section*{Methods}
\subsection*{Electronic density-of-states fingerprints \label{sec:dos-fingerprint}}
The analysis of spectra like the DOS is typically done by visual inspection, \ie in a qualitative manner. For large data sets, this kind of analysis quickly becomes unfeasible. Therefore, a descriptor that allows for automated processing of such data is required. This includes a suitable numerical representation of the DOS. In the following, we review such representations that have been proposed in the literature, state their drawbacks, and tell how we overcome them.

To quantitatively compare the DOS of materials, Isayev \textit{et al.}\cite{Curtarolo1} constructed a DOS \textit{fingerprint} by encoding the DOS in the energy range between $-10$ and $10$ $\mathrm{eV}$ as a series of $256$ float (4 bytes) numbers. A similar point-wise representation was considered in Ref. [\cite{learningDOS}] for building predictive models for the DOS based on Gaussian regression. 
It was pointed out that such a representation were inefficient, as it may potentially require many sampling points of the DOS to efficiently train the models. Moreover, loss functions based on that representation turned out largely insensitive to spectral features with small overlap. To overcome these problems, the authors proposed two approaches: \textit{i)} a truncated basis expansion based on principal-component analysis (PCA), which leads to an effective reduction of the degrees of freedom of the fingerprint (effectively smoothing the DOS spectra), and \textit{ii)} a representation based on the cumulative distribution function associated to the DOS. The latter improved on the sensitivity of the loss functions to non-overlapping spectral features. More recently, a high-dimensional fingerprint based on the DOS projected on different atomic orbitals and sites, followed by PCA dimensionality reduction, has been proposed \cite{C2DBprogress}.

A common drawback of all these DOS fingerprints is that they  equally weigh the contributions from the entire energy range considered in the spectra. Thus, they don't account for the fact that different energy regions are associated to distinct physical phenomena. For instance, the shape of the DOS close to the top of the valence band and the size of the band-gap are most important aspects in the search for p-doped materials. Likewise, for metals, the magnitude and shape of the DOS around the Fermi energy are most relevant. Other research may focus on some features of the conduction band. Although the PCA-based approach mentioned above can effectively lead to a re-weighting of spectral features,  
it cannot be tailored at will to focus on specific regions, but is determined by the training data that is used for the construction of the descriptors.

\begin{figure}[tb]
    \centering
    \includegraphics[width = 0.45\textwidth]{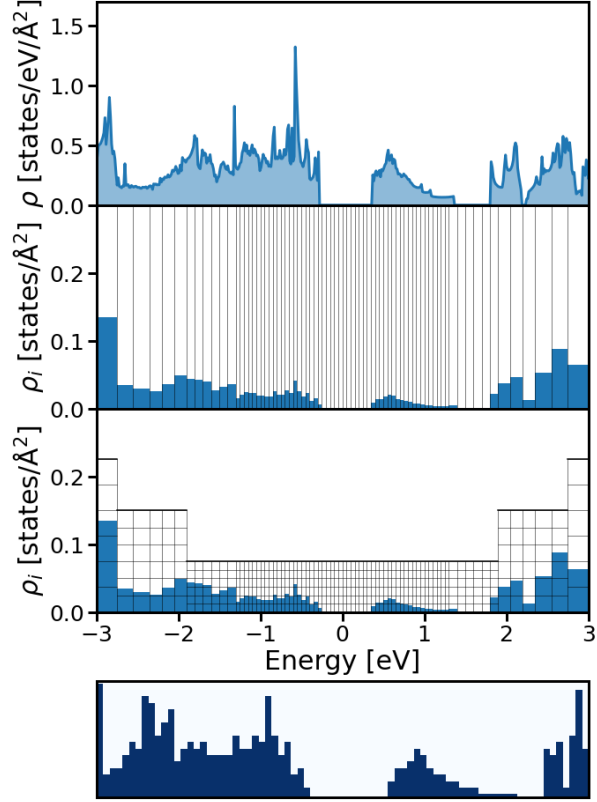}
    \caption{ Generation of DOS fingerprints (bottom panel) from the electronic DOS, $\rho(E)$ (top panel). The DOS of a material (top panel) is numerically integrated over small energy intervals $[\varepsilon_i,\varepsilon_i+\Delta\varepsilon_i)$ (Eq. \ref{eq:dos_bins}). The thereby generated histogram of states (second panel) is subsequently discretized (third panel), resulting in an image (bottom panel) of the DOS. 
    In this image, each dark (light) pixel  corresponds to a 1 (0) in the fingerprint. In the third panel, only every fifth discretization step is shown. To increase visibility, we use $N_\rho = 30$, $\rho_\mathrm{min} = 0.075$, and $\rho_\mathrm{max} = 0.825$ for this figure. The other parameters are set as described in the Code Availability section. 
    \label{fig:dos_fp_gen}}
\end{figure}

To overcome the described issues, we have developed a DOS fingerprint that allows for a tailored weighting of spectral features. Using a non-uniform discretization of the energy axis, the fingerprint can be adapted to focus on desired energy regions. 
To achieve this discretization, the DOS is transformed into a two-dimensional raster image (Fig. \ref{fig:dos_fp_gen}, bottom panel) as follows:  First, the spectrum is shifted such that the energy $\varepsilon = 0$ is located at a reference energy $\varepsilon_\mathrm{ref}$, which defines the main focus of the fingerprint. 
Then, the DOS ($\rho(E)$,  Fig. \ref{fig:dos_fp_gen} top) is integrated over an even number $N_\varepsilon$ of intervals of variable widths $\Delta\varepsilon_i$, to obtain a histogram $\{\rho_i\}$ (Fig. \ref{fig:dos_fp_gen}, second panel):
\begin{eqnarray}
\rho_i = \int_{\varepsilon_i}^{\varepsilon_{i+1}} \rho(\varepsilon) d\varepsilon,
\label{eq:dos_bins}
\end{eqnarray}
with $i\in\left[-N_\varepsilon/2,\,N_\varepsilon/2\right]$, $i \in \mathbb{Z}$, $\varepsilon_0 = 0$, $\varepsilon_{i+1}=\varepsilon_{i}+\Delta \varepsilon_i$ for $i\ge 0$, and $\varepsilon_{-i} = -\varepsilon_i$. 
The integration intervals  $\Delta \varepsilon_i$ are defined as
\begin{eqnarray}
\Delta \varepsilon_i = n(\varepsilon_i,W,N)\, \Delta \varepsilon_{min},\label{eq:eps_i}
\end{eqnarray}
where $\Delta \varepsilon_{min}$ is a parameter giving the minimal integration width and the integer-valued function
\begin{eqnarray}
 n(\varepsilon,W,N) = \lfloor g(\varepsilon,W)  N+1 \rfloor \in \left[1,N\right],\label{eq:f_grid}
\end{eqnarray}
where $\lfloor \cdot \rfloor$ denotes the 'round down' operator and $g(\varepsilon,W) = (1 - \exp(-\varepsilon^2/2 W^2))$. Here, the parameter $N\in \mathbb{N}$ ($N>1$) determines the maximum interval width $N\Delta \varepsilon_{min}$, and the parameter $W$ determines the \textit{feature region}: For $\varepsilon = 0$, $\Delta \varepsilon_i $ equals $\Delta \varepsilon_{min}$, while it approaches $N\Delta \varepsilon_{min}$ for $|\varepsilon|> W$. In this way, a finer discretization of the histogram is obtained for energies in the feature region $|\varepsilon|<W$. This is illustrated by the integration limits indicated by vertical lines in the second panel of Fig. \ref{fig:dos_fp_gen}. From this histogram, a raster graphic is generated by defining a grid of pixels, as shown in the third panel of Fig. ~\ref{fig:dos_fp_gen}. Every column $i$ of the histogram is discretized in a grid of $N_\rho$ intervals of height
\begin{eqnarray}
\Delta \rho_{i} =  n(\varepsilon_i,W_H, N_H)\,\Delta \rho_{min}. \label{eq:rho_grid}
\end{eqnarray}
 
Here, the parameters $W_H$, $N_H$, and $\Delta \rho_{min}$ play a role analogous to $W$, $N$, and $\Delta \varepsilon_{min}$ above: Close to $\varepsilon=0$, a fine discretization $\Delta \rho_{i}=\Delta \rho_{min}$ is obtained, while it approaches $N_H\Delta \rho_{min}$ for $|\varepsilon|>W_H$. Finally, the number of "filled" pixels in column $i$ is determined by
\begin{eqnarray}
\min\left(\left\lfloor\frac{\rho_i}{\Delta \rho_{i}}\right\rfloor,N_\rho\right),
\end{eqnarray}
resulting in the 2D raster image of the bottom panel of Fig.~\ref{fig:dos_fp_gen}, containing $N_\varepsilon\times N_\rho$ pixels enumerated by an index $\alpha$. This image is then transformed into a binary-encoded vector $\bm{f}=(f_1,...,f_{N_\varepsilon\times N_\rho})$ with component $f_\alpha=1$ if the pixel $\alpha$ is filled and $0$ otherwise. 

\subsection*{DOS similarity metric}

The similarity between two materials $i$ and $j$ in terms of their DOS fingerprints $\bm{f}_i$ and $\bm{f}_j$ is denoted by $S(\bm{f}_i,\bm{f}_j)$. As similarity metric, we use the Tanimoto coefficient (Tc) \cite{similarity_measures}, defined as:
\begin{eqnarray}
S(\bm{f}_i, \bm{f}_j) = \frac{\bm{f}_i \cdot \bm{f}_j}{|\bm{f}_i|^2 + |\bm{f}_j|^2 - \bm{f}_i \cdot \bm{f}_i}. \label{eq:similarity_measure}
\end{eqnarray}
$S(\bm{f}_i, \bm{f}_j)$ can be interpreted as the overlap of the areas covered by the raster images represented by $\bm{f}_i$ and $\bm{f}_j$, divided by the union of the areas. $S$ takes real values in the range $[0,1]$, being equal to $1$ ($0$) if the images $\bm{f}_i$ and $\bm{f}_j$ are identical (have no overlap). 

\begin{figure}[tb]
    \centering
    \includegraphics[width = 0.6\linewidth]{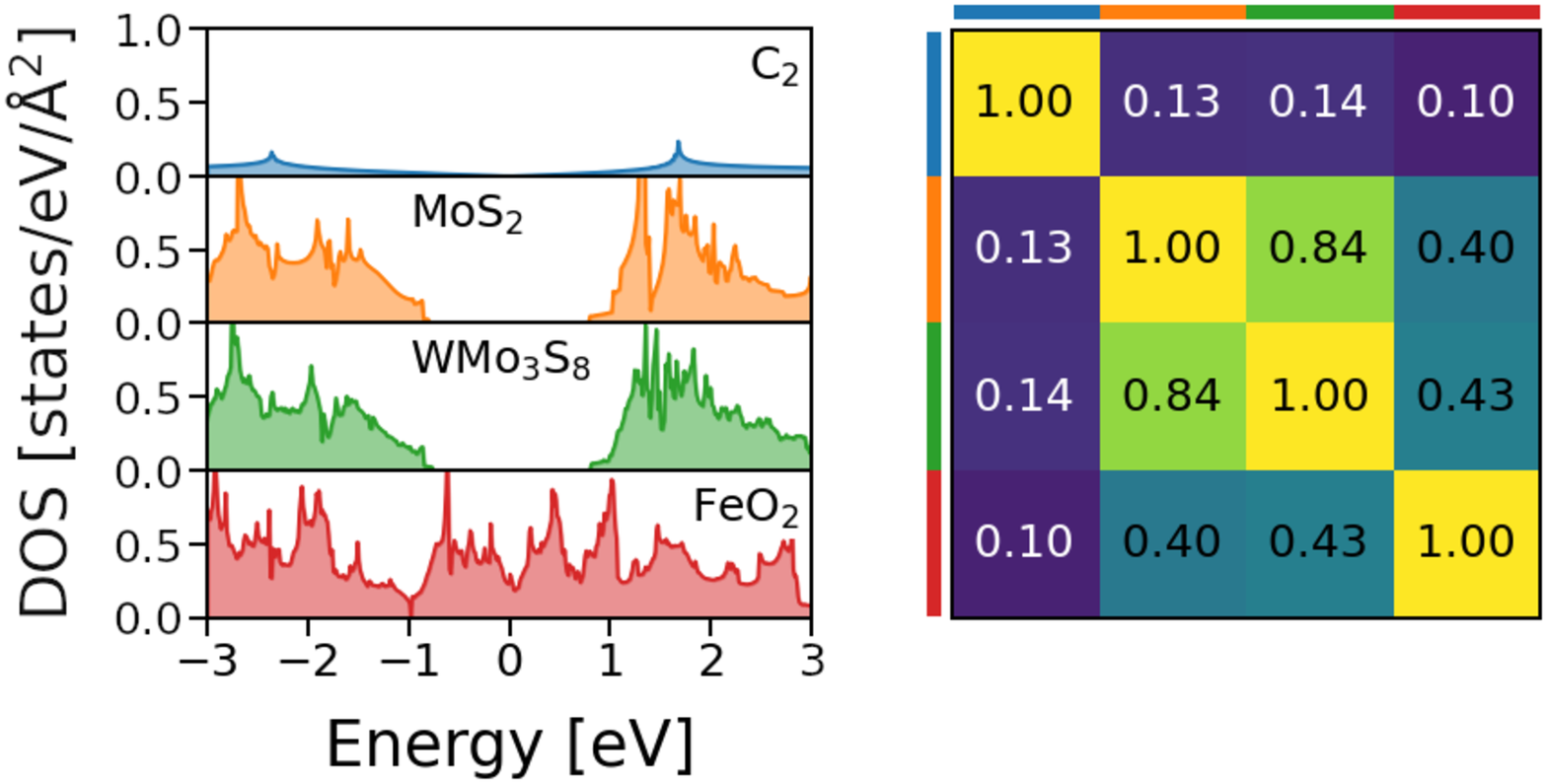}
    \caption{Illustration of the similarity metric with the examples of four materials. The DOS of graphene (C$_2$), MoS\fsub{2}, WMo\fsub{3}S\fsub{8}, and FeO\fsub{2} are presented on the left. The Fermi level is located at $E = 0$ eV. The similarity matrix (rows and columns in the same order, color-coded) is shown in the right panel.}
\label{fig:four_dos}
\end{figure}
As an example of this metric, Fig.~\ref{fig:four_dos} shows the DOS of four different materials from the C2DB and their respective similarities. In the considered energy interval, C\fsub{2} (graphene) has much fewer available states than the other examples. Mainly for this reason, it is dissimilar to all of them ($S\le 0.14$, see similarity matrix in Fig.~{\ref{fig:four_dos}}). The DOS of MoS\fsub{2} is similar to that of FeO\fsub{2} in magnitude for $|\varepsilon|>1 eV$, but since MoS\fsub{2} is a semiconductor and FeO\fsub{2} is a metal, the overall similarity is low ($S=0.4$). MoS\fsub{2} and WMo\fsub{3}S\fsub{8} exhibit a high similarity coefficient of $S=0.84$, as both the shape and the magnitude are similar.

\subsection*{Additional similarity measures} \label{sec:add_similarity_measures}
In this work, we cluster materials based on the DOS similarity fingerprint (see Methods section). To understand why such clusters are formed, we make use of additional descriptors. 

The electronic spectrum of a material can often be understood by counting the valence electrons in the outermost shells of its constituent atoms. This counting can, in principle, be obtained from the average of
the column numbers of the atoms in the unit cell:
\begin{eqnarray}
\overline{c}_m = \frac{1}{N} \sum_{i}^{N} c_{im} \label{eq:average_column_descriptor} 
\end{eqnarray}
where $i$ runs over all $N$ atoms in the unit cell of material $m$, and $c_{im}$ denotes their column in the PTE. $\overline{c}_m$ is calculated for all materials in a cluster. If it is equal for all of them, we conclude that the cluster is formed by isoelectronic materials. Note that here we employ a lax definition of isoelectronicity that considers only electron counting but not electronic configuration. As an example, $\overline{c}_m$ for two Si atoms is identical that of two C atoms or the combination of one Al and one P atom.
We call this descriptor the {\it PTE descriptor}.

The geometry of the crystal structures can also be explicatory of clusters obtained from the DOS similarity metric. Accordingly, we consider a similarity measure based on the space group (SG) of the crystal structures, after removing all information of the species that form the structure.  In practice, this is achieved by first replacing all atoms by a single species and then employing the software package \texttt{spglib} \cite{spglib} with a tolerance of $\texttt{symprec}=1\times10^{-1}$ to find the SG of the resulting geometry. In the main text we call this the {\it SG descriptor}. 

\subsection*{Clustering algorithm} \label{sec:clustering}
A similarity metric allows for a range of practical applications as, for instance, to determine which materials from a data set are most similar to any given reference. 
The latter could be a material with a desired property, for which one seeks alternatives. 
This kind of analysis is commonly applied in chemical similarity searching \cite{similarity_measures, medicinal_chemistry} or drug discovery \cite{molecular_similarity}. A related application is the detection of (sub)sets of materials, \ie clusters, that are more similar to one another than to other materials. In this work, we focus on the second case and develop a clustering algorithm that takes advantage of the following property of our similarity measure (Eq. \ref{eq:similarity_measure}): Its complement $1 - \mathrm{S}$ is a distance measure that is identical to the Soergel distance for dichotomous fingerprints \cite{similarity_measures}. For binary-valued descriptors, it obeys the triangle inequality\cite{similarity_measures}, \ie
\begin{eqnarray}
S(\bm{f}_i, \bm{f}_j) \geq S(\bm{f}_i, \bm{f}_k) + S(\bm{f}_k, \bm{f}_j) - 1 \label{eq:triangle_inequality}
\end{eqnarray}
for any three fingerprints $\bm{f}_i, \bm{f}_j,$ and $ \bm{f}_k$. This can be easily verified with the examples shown in Fig.~\ref{fig:four_dos}.
An important consequence is that any two materials that are more similar to a third one than a threshold $S_\mathrm{thres}$, will be more similar than $2 S_\mathrm{thres} - 1$ to each other. For example, if we choose $S_\mathrm{thres}(\bm{f}_k, \bm{f}_{\rm{ref}}) = 0.75$, all materials $\bm{f}_k$ within a cluster centered at the reference material $\bm{f}_{\rm{ref}}$ have $S \geq 0.5$ to all other cluster members. This motivates a simple clustering algorithm as follows: Start by \textit{i)} making a list of the materials in the database. Then, \textit{ii)} identify the material (reference) with the highest number of other materials that are more similar to it than a given threshold $S_\mathrm{thres}$. If no materials can be found for any reference, stop the algorithm, as all possible clusters are found. Otherwise, \textit{iii)} consider the found reference and its similar materials as a cluster and extract them from the list; and return to step \textit{ii)}. The materials that do not belong to any cluster are considered \textit{orphans}. When two materials have the same number of neighbors and share any of them, the cluster with the highest average similarity is selected. 

\section*{Code availability} \label{sec:comp_params}

Our implementation of the DOS fingerprint is part of the NOMAD project \cite{NOMAD2} and can be obtained as a stand-alone Python package from Ref.~[\cite{dos_fingerprints}]. For our analysis, we choose to focus our DOS descriptor on an energy region around the Fermi energy.
To this end, we set the parameters $\Delta \varepsilon_{min} = 0.05$ eV, $\Delta \varepsilon_{max} = 1.05$ eV, $N=\Delta \varepsilon_{max}/\Delta \varepsilon_{min}=21$, $\varepsilon_\mathrm{ref} = 0$ eV, and $W = 4$ eV (see Eq.~\ref{eq:eps_i}). 
 $W_H = 4$ eV, $N_{\rho} = 512$, $\rho_{min} = N_{\rho}\Delta \rho_{min}=0.25$, $\rho_{max} = 2.75$, and $N_H=\rho_{max}/\rho_{min}=11$ (see Eq. \ref{eq:rho_grid}). An implementation of our clustering algorithm is available at Github [\cite{clustering_algorithm}].

\section*{Data preprocessing}
\label{sec:data}
We use data from the Computational 2D Materials Database (C2DB) \cite{Haastrup_2018, C2DBprogress}, a high-throughput database of 2D materials. The majority of its content is generated from structure prototypes that are decorated with different atomic species. At the point of writing this manuscript, it contained 4047 structures, composed of 63 different chemical elements. Projected densities of states (PDOS) and atomic structures are available at the C2DB website  \cite{c2dburl} for 3491 of these structures. The C2DB contains various materials properties calculated by the density-functional-theory (DFT) package GPAW \cite{Mortensen2005,Enkovaara2010}, employing the generalized-gradient approximation (GGA) in the PBE parameterization. The data were generated in an automated manner using the Atomic Simulation Recipe (ASR) \cite{asr} package. 

Before using this data for our analysis, we preprocess it in the following way: First, we sum up over all PDOS for one material, in order to obtain the total DOS (TDOS). Then, we define the Fermi level, $E_F$, as the energy zero. The TDOS of every structure is then normalized with respect to the area of the unit cell spanned by the two periodic cell vectors. In this way, the results can be consistently compared across different geometries. For instance, supercells of the same material are considered identical. The resulting normalized TDOS are then employed to generate the fingerprints that encode the electronic structure. Analysis of atomic structures are performed using the Python package \texttt{ASE} \cite{ASE}. Further plots are generated using \texttt{matplotlib} \cite{matplotlib}. 

\section*{Data availability}

The data used in this publication can be accessed through a public web application programming interface (API)\cite{c2dburl}. 

\bibliographystyle{unsrt}
\bibliography{main}

\section*{Acknowledgements}

This work received partial funding by the German Research Foundation (DFG) through the CRC 1404 (FONDA), Projektnummer 414984028, and the NFDI consortium FAIRmat – project 460197019. Partial support from the the European Union’s Horizon 2020 research and innovation program under the grant agreement Nº 951786 (NOMAD CoE) is appreciated.

\section*{Author contributions statement}

M.K. prepared the data, wrote the code, and analyzed the results, M.S. contributed to the development of the fingerprint, S.R. and C.D. supervised and reviewed all parts of the work. All authors contributed to the writing of the manuscript.

\section*{Competing interests}

The authors declare no competing interests.

\end{document}